\pdfoutput=1
\RequirePackage{ifpdf}
\ifpdf % We are running pdfTeX in pdf mode
\documentclass[pdftex]{sigma}
\else
\documentclass{sigma}
\fi

\newcommand{\starco}[2]{[#1\stackrel{\star}{,}#2]}

\DeclareMathOperator{\const}{const}
\DeclareMathOperator{\diag}{diag}
\DeclareMathOperator{\Tr}{Tr}
						
\numberwithin{equation}{section}

\begin{document}

\newcommand{\arXivNumber}{1403.0255}

\allowdisplaybreaks

\renewcommand{\thefootnote}{$\star$}

\renewcommand{\PaperNumber}{099}

\FirstPageHeading

\ShortArticleName{Wong's Equations and Charged Relativistic Particles
in Non-Commutative Space}

\ArticleName{Wong's Equations and Charged Relativistic Particles\\ in Non-Commutative Space\footnote{This paper is a~contribution
to the Special Issue on Deformations of Space-Time and its Symmetries.
The full collection is available at \href{http://www.emis.de/journals/SIGMA/space-time.html}
{http://www.emis.de/journals/SIGMA/space-time.html}}}

\Author{Herbert BALASIN~$^\dag$, Daniel N.~BLASCHKE~$^\ddag$, Fran\c{c}ois GIERES~$^\S$
and Manfred SCHWEDA~$^\dag$}

\AuthorNameForHeading{H.~Balasin, D.N.~Blaschke, F.~Gieres and M.~Schweda}

\Address{$^\dag$~Institute for Theoretical Physics, Vienna University of Technology,\\
\hphantom{$^\dag$}~Wiedner Hauptstra\ss e 8-10, A-1040 Vienna, Austria}
\EmailD{\href{mailto:hbalasin@tph.tuwien.ac.at}{hbalasin@tph.tuwien.ac.at},
\href{mailto:mschweda@tph.tuwien.ac.at}{mschweda@tph.tuwien.ac.at}}

\Address{$^\ddag$~Los Alamos National Laboratory, Theory Division, Los Alamos, NM, 87545, USA}
\EmailD{\href{mailto:dblaschke@lanl.gov}{dblaschke@lanl.gov}}

\Address{$\S$~Universit\'e de Lyon, Universit\'e Claude Bernard Lyon 1 and CNRS/IN2P3,\\
\hphantom{$^\S$}~Institut de Physique Nucl\'eaire, Bat.~P.~Dirac,\\
\hphantom{$^\S$}~4 rue Enrico Fermi, F-69622-Villeurbanne, France}
\EmailD{\href{mailto:gieres@ipnl.in2p3.fr}{gieres@ipnl.in2p3.fr}}

\ArticleDates{Received March 02, 2014, in f\/inal form October 17, 2014; Published online October 24, 2014}

\Abstract{In analogy to Wong's equations describing the motion of a~charged relativistic point particle in the presence of an
external Yang--Mills f\/ield, we discuss the motion of such a~particle in non-commutative space subject to an external $U_\star(1)$
gauge f\/ield.
We conclude that the latter equations are only consistent in the case of a~constant f\/ield strength.
This formulation, which is based on an action written in Moyal space, provides a~coarser level of description than full QED on
non-commutative space.
The results are compared with those obtained from the dif\/ferent Hamiltonian approaches.
Furthermore, a~continuum version for Wong's equations and for the motion of a~particle in non-commutative space is derived.}


\Classification{81T13; 81T75; 70S05}

\renewcommand{\thefootnote}{\arabic{footnote}} \setcounter{footnote}{0}

\section{Introduction}

Over the last twenty years, non-commuting spatial coordinates have appeared in various contexts in the framework of quantum
gravity and superstring theories.
This fact contributed to the motivation for studying classical and quantum theories with a~f\/inite or inf\/inite number of degrees
of freedom on non-commutative spaces.
Dif\/ferent mathematical approaches have been pursued and various physical applications have been explored, e.g.\
see references~\cite{Blaschke:2010kw,Delduc:2007av,Horvathy:2010wv,Rivasseau:2007a, Szabo:2001} for some partial reviews.
Beyond the applications, classical and quantum mechanics on non-commutative space are of interest as toy models for f\/ield
theories which are more dif\/f\/icult to handle, in particular in the case of interactions with gauge f\/ields.
In this respect, we recall a~similar situation concerning the coupling of matter to Yang--Mills f\/ields on ordinary space:
A~coarser level of description for the latter theories has been proposed by Wong~\cite{Wong:1970fu} who considered the motion of
charged point particles in an external gauge
f\/ield~\cite{Balachandran:1977ub,Balachandran:1983pc,Brown:1979bv,Duval:1980uk,Duval:1981js,Kosyakov:1998qi,Kosyakov:2007qc,Wipf:1985ak}.
The latter equations allow for various physical applications, e.g.\
to the dynamics of quarks and their interaction with gluons~\cite{Kosyakov:1998qi, Kosyakov:2007qc}.
Somewhat similar equations, known as Mathisson--Papapetrou--Dixon equations~\cite{Dixon:1970zza, Mathisson:1937zz,Papapetrou:1951pa}
appear in general relativity for a~spinning particle in curved space.
In this spirit, we will consider in the present work the dynamics of a~relativistic ``point'' particle in non-commutative space
subject to an external $U_\star(1)$ gauge f\/ield (thereby implementing a~suggestion made in an earlier work~\cite{Stern:2007an},
see also~\cite{Fatollahi:2004nt} for some related studies based on a~f\/irst order expansion in the non-commutativity parameter).

More precisely, we are motivated by the motion in Moyal space, the latter space having been widely discussed as the arena for
f\/ield theory in non-commutative spaces.
By proceeding along the lines of Wong's equations (which are discussed in~Section~\ref{sec:recallcommspace}), we derive a~set of
equations for the dynamics of the particle in Moyal space in Section~\ref{sec:lagrange}.
For the description of the coupling of a~particle to a~gauge f\/ield, the relativistic setting is the most natural one, but our
discussions (of Wong's equations in commutative space or of a~particle in non-commutative space) could equally well be done
within the non-relativistic setting.
The particular case of a~constant f\/ield strength is the most tractable one (and the only consistent one for Wong's equations in
non-commutative space) and will be considered in some detail in~Section~\ref{sec:constfield}.
Subsequently in~Section~\ref{sec:ham}, we brief\/ly recall the dif\/ferent Hamiltonian approaches which have previously been pursued
for the formulation of classical mechanics in non-commutative space and we compare the resulting equations governing the dynamics
of particles coupled to an electromagnetic f\/ield.
In the appendix we present a~continuum formulation of Wong's equations on a~generic space-time manifold, a~formulation which readily
generalizes to Moyal space.

For the motion of a~relativistic point particle in four-dimensional (commutative or non-commutative) space-time, the following
notations will be used.
The metric tensor is given~by
\begin{gather*}
(\eta_{\mu \nu})_{\mu, \nu \in \{0, 1,2,3 \}}= \diag(1, - 1, -1, -1)
\end{gather*}
and we choose the natural system of units ($c\equiv 1 \equiv \hbar$).
The proper time~$\tau$ for the particle is def\/ined (up to an additive constant)~by
\begin{gather}
\label{defpt}
d\tau^2= ds^2,
\qquad
\text{with}
\qquad
ds^2 \equiv dx^{\mu} dx_{\mu}= (dt)^2 - (d\vec x)^2,
\end{gather}
and for the massive particle we have $ds^2 > 0$.
From~\eqref{defpt} it follows that $\dot{x}^2=1$ where $\dot{x}^2 \equiv \dot{x}^{\mu} \dot{x}_{\mu} $ and $ \dot{x}^{\mu}
\equiv {dx^{\mu}}/{d\tau} $.

\section{Reminder on particles in commutative space}
\label{sec:recallcommspace}

{\bf Abelian gauge f\/ield in f\/lat space.} We consider the interaction of a~charged massive relativistic particle with an external
electromagnetic f\/ield given by the $U(1)$ gauge potential $(A^\mu)$.
The motion of this particle along its space-time trajectory $\tau \mapsto x^\mu(\tau)$ is described by the action\footnote{To be
more precise, in the integral~\eqref{eq:point-action-comm-u1} the variable~$\tau$ is viewed as a~purely mathematical parameter
which is only identif\/ied with proper time after deriving the equations of motion from the action.
Thus, the relation $\dot{x}^2=1$ is only to be used at the latter stage.}
\begin{gather}
S[x]=-m\int ds-q\int dx^\mu A_\mu(x(\tau))
= -m \int d\tau\sqrt{\dot{x}^2} -q\int d\tau\dot x^\mu A_\mu(x(\tau)).
\label{eq:point-action-comm-u1}
\end{gather}
Here,~$q$ denotes the conserved electric charge of the particle associated with the conserved current density $(j^\mu)$:
\begin{gather}
j^\mu(y)=q\int d\tau\dot x^\mu(\tau)\delta^4(y-x(\tau)),
\qquad
\partial_\mu j^\mu=0,
\qquad
\int d^3yj^0(y)=q.
\label{eq:abelian-current}
\end{gather}
We note that the interaction term in the functional~\eqref{eq:point-action-comm-u1} may be rewritten in terms of the above
current according to
\begin{gather}
\label{jA}
q\int d\tau\dot x^\mu A_\mu(x(\tau))=\int d^4yj^\mu(y)A_\mu(y).
\end{gather}
This expression is invariant under inf\/initesimal gauge transformations, i.e.~$\delta_\lambda A_\mu= \partial_\mu \lambda$,
thanks to the conservation of the current:
\begin{gather*}
\delta_\lambda\int d^4yj^\mu A_\mu=\int d^4yj^\mu \partial_\mu\lambda=-\int d^4y(\partial_\mu j^\mu) \lambda=0.
\end{gather*}
We note that the parameter~$q$ which describes the coupling of the particle to the gauge f\/ield might in principle depend on the
world line parameter~$\tau$: if this assumption is made, $q(\tau)$ appears under the~$\tau$-integral
in~\eqref{eq:abelian-current}, the current $j^\mu$ is no longer conserved and the coupling~\eqref{jA} is no longer gauge
invariant.
Thus,~$q$ is necessarily constant along the path.

Variation of the action~\eqref{eq:point-action-comm-u1} with respect to $x^\mu$ and substitution of the relation $\dot{x}^2=1$
leads to the familiar equation of motion
\begin{gather*}
m\ddot x^\mu=q F^{\mu\nu}\dot x_\nu,
\qquad
\text{with}
\qquad
F_{\mu \nu} \equiv \partial_\mu A_\nu - \partial_\nu A_\mu.
\end{gather*}
As \looseness=-1 is well known, the free particle Lagrangian $L_{\rm free}= -m \sqrt{\dot{x}^2}$
which represents the f\/irst term of the
action~\eqref{eq:point-action-comm-u1} and which is non-linear in $\dot{x}^2$ can be replaced by $\tilde{L}_{\rm free}=
\frac{m}{2}\dot{x}^2 $ which is linear in~$\dot{x}^2$ since both Lagrangians yield the same equation of motion.
Indeed, we will consider the latter Lagrangian in equation~\eqref{eq:LGrass} and in~Section~\ref{sec:ham} where we go over to the
Hamiltonian formulation.

The aim of this work is to generalize this setting to a~non-commutative space-time.
Since a~gauge f\/ield on non-commutative space entails a~non-Abelian structure for the f\/ield strength tensor $F_{\mu\nu}$ due to
the star product, it is worthwhile to understand f\/irst the coupling of a~spinless particle to a~non-Abelian gauge f\/ield on
commutative space-time.

{\bf Non-Abelian gauge f\/ield in f\/lat space.} We consider a~compact Lie group~$G$ (e.g.\
$G=\text{SU}(N)$ for concreteness) with generators $T^a$ satisfying
\begin{gather*}
\big[T^a, T^b\big]= {\rm i} f^{abc} T^c
\qquad
\text{and}
\qquad
\Tr\big(T^a T^b\big)=\delta^{ab}.
\end{gather*}
Just as for the Abelian gauge f\/ield, the source $j_\mu^a (x)$ of the non-Abelian gauge f\/ield $A_{\mu}(x) \equiv A_{\mu}^a (x)
T^a$ (e.g.\
a~f\/ield theoretic expression like $\bar{\psi} \gamma_\mu \hat{T}^a \psi$ involving a~multiplet~$\psi$ of spinor f\/ields) is
considered to be given by the current density $j_\mu^a (x)$ of a~relativistic point
particle~\cite{Balachandran:1977ub,Balachandran:1983pc,Duval:1980uk,Duval:1981js,Kosyakov:1998qi,Kosyakov:2007qc,Wipf:1985ak, Wong:1970fu}.
Instead of an electric charge, the particle moving in an external Yang--Mills f\/ield is thus assumed to carry a~color-charge or
isotopic spin $\vec q \equiv (q^a)_{a=1, \dots, \text{dim}G}$ which transforms under the adjoint representation of the
structure group~$G$.
Henceforth, one considers the Lie algebra-valued variable $q(\tau) \equiv q^a(\tau) T^a$ which is assumed to be~$\tau$-dependent.
The particle is then described in terms of its space-time coordinates $x^\mu (\tau)$ and its isotopic spin $q(\tau)$, i.e.~it is
referred to with respect to geometric space and to internal space.
For the moment, we assume~$q(\tau)$ to represent a~given non-dynamical (auxiliary) variable and we will comment on a~dif\/ferent
point of view below.
Its coupling to an external non-Abelian gauge f\/ield $(A_{\mu})$ is now described by the action
\begin{gather}
S[x]=-m\int ds-\int d\tau \dot x^\mu\Tr \{q(\tau)A_\mu(x(\tau)) \}
\nonumber
\\
\phantom{S[x]}
= -m \int d\tau\sqrt{\dot{x}^2} -\int d\tau\dot x^\mu q^a(\tau)A^a_\mu(x(\tau)).
\label{eq:point-action-comm-un}
\end{gather}
The current density~\eqref{eq:abelian-current} presently generalizes to a~Lie algebra-valued expression $j_\mu \equiv j_{\mu}^a
T^a$ given~by
\begin{gather}
j^\mu(y)=\int d\tau\dot x^\mu(\tau)q(\tau)\delta^4(y-x(\tau)),
\label{JNA}
\end{gather}
and thereby the interaction term in the functional~\eqref{eq:point-action-comm-un} can be rewritten as
\begin{gather}
- \int d^4y\Tr(j^\mu A_\mu).
\label{JAcoupling}
\end{gather}
For an inf\/initesimal gauge transformation with Lie algebra-valued parameter~$\lambda$, i.e.~$\delta_{\lambda}A_\mu= D_\mu
\lambda \equiv \partial_\mu \lambda -{\rm i} g[A_\mu, \lambda]$, we have
\begin{gather}
\delta_\lambda\int d^4y\Tr\{j^\mu A_\mu \}=\int d^4y\Tr\{j^\mu D_\mu\lambda \}=-\int d^4y\Tr\{(D_\mu j^\mu) \lambda \}.
\end{gather}
Thus, gauge invariance of the action~\eqref{eq:point-action-comm-un} requires the current to be \emph{covariantly conserved,}
i.e.~$D_\mu j^\mu=0$.
From~\eqref{JNA} we can deduce by a~short calculation that
\begin{gather}
(D_\mu j^\mu)^a (y)=\int d\tau \frac{D q^a}{d\tau}\delta^4(y-x(\tau)),
\qquad
\text{with}
\qquad
\frac{D q^a}{d\tau} \equiv \frac{d q^a}{d\tau} -{\rm i} g \dot{x}^{\mu} [A_\mu (x(\tau)),q]^a,
\label{DJNA}
\end{gather}
hence $j^\mu$ is covariantly conserved if the charge~$q$ is \emph{covariantly constant} along the world line: $ {D q^a}/{d\tau}
=0 $ (subsidiary condition).

Variation of the action with respect to $x^\mu$ (and use of $\dot{x}^2=1$) yields the equations of motion
\begin{gather}
\label{wong}
m\ddot x^\mu= \Tr(q F^{\mu\nu})\dot x_\nu,
\qquad
\text{where}
\qquad
\frac{D q^a}{d\tau}=0,
\end{gather}
and $F_{\mu \nu} \equiv \partial_\mu A_\nu - \partial_\nu A_\mu - {\rm i} g[A_\mu, A_\nu] $.
By construction, these equations are invariant under inf\/initesimal gauge transformations parametrized by $\lambda (x)$ since the
charge~$q$ is a~non-dynamical (auxiliary) variable transforming as $ \delta_{\lambda} q (\tau) \equiv -{\rm i}g[q(\tau), \lambda
(x (\tau))]=0$ for all~$\tau$, hence
\begin{gather*}
 \delta_{\lambda} \big(\Tr(q F_{\mu\nu})\dot x^{\nu} \big)= -{\rm i}g\Tr(q[F_{\mu\nu},
\lambda])\dot x^{\nu}={\rm i}g\Tr([q, \lambda]F_{\mu\nu})\dot x^{\nu}=0,
\\
 \delta_{\lambda} \left(\frac{D q}{d\tau} \right)= -{\rm i}g\dot x^{\mu}[\delta_{\lambda} A_{\mu}, q]
= -{\rm i}g\left[\frac{D \lambda}{d\tau}, q \right]= -{\rm i}g
\frac{D}{d\tau}[\lambda, q]=0.
\end{gather*}
Equations~\eqref{wong}, which represent the Lorentz--Yang--Mills force equation for a~relativistic particle in an external
Yang--Mills f\/ield, are known as \emph{Wong's equations}~\cite{Wong:1970fu}.
More specif\/ically, the relation $Dq^a /d\tau=0$ may be viewed as charge transport equation and it is the geometrically natural
generalization (to the charge vector $(q^a)$) of the constancy of the charge~$q$ in electrodynamics.

A few remarks are in order here.
We should point out that the equation of motion for $x^{\mu}$ had already been obtained earlier in curved space~by
Kerner~\cite{Kerner:1968fn}, but the charge transport equation was not established in that setting.
We refer to the works~\cite{Balachandran:1977ub,Balachandran:1983pc,Kosyakov:1998qi,Kosyakov:2007qc} for a~treatment of dynamics
involving the Lagrangian $-\frac{1}{4} \int d^3 x\Tr(F^{\mu \nu} F_{\mu \nu})$ of the gauge f\/ield: the latter
approach yields the covariant conservation law $D_{\mu} j^{\mu}=0$ as a~consequence of the equation of motion $D_{\nu} F^{\nu
\mu}= j^{\mu}$ and the relation $[D_{\mu}, D_{\nu}] F^{\nu \mu}=0$.
For a~general discussion of the issue of gauge invariance for the coupling of gauge f\/ields to non-dynamical external sources, we
refer to~\cite{Ramond:1981pw}.
Equations~\eqref{wong} and their classical solutions have been investigated in the literature and applied for instance to the
study of the quark gluon plasma~\cite{Kosyakov:1998qi}.
The particular case of a~constant f\/ield strength $F_{\mu \nu}$ (``uniform f\/ields'') exhibits interesting
features~\cite{Brown:1979bv} which will be commented upon in~Section~\ref{sec:constfield}.

If one regards $(q^a)$ as a~dynamical variable which satisf\/ies the equation of motion ${D q^a}/{d\tau}=0$ and which transforms
under gauge variations with the adjoint representation, i.e.~
\begin{gather}
\label{eq:transfq}
\delta_{\lambda} q(\tau)= -{\rm i}g[q(\tau), \lambda (x (\tau))],
\end{gather}
then equations~\eqref{wong} are obviously gauge invariant since $F_{\mu \nu}$ also transforms with the adjoint
representation\footnote{If $j^\mu$ is assumed to transform covariantly, then gauge invariance of the action~\eqref{JAcoupling}
implies $\partial_\mu j^\mu=0$ as has already been noticed in reference~\cite{Sikivie:1978sa}.}.
However, we emphasize that we started from the action~\eqref{eq:point-action-comm-un} to obtain the equation of motion of~$x^\mu$, the one of~$q$ following from the requirement of gauge invariance of the initial action\footnote{In fact, if the
charge~$q$ is treated as a~dynamical variable in the action~\eqref{eq:point-action-comm-un}, then it amounts to a~Lagrange
multiplier leading to the equation of motion $\dot{x}^\mu A^a_\mu(x(\tau))=0$ which is not gauge invariant.
We thank the anonymous referee for drawing our attention to this point.}.
An action which yields the equations~\eqref{wong} as equations of motion of both~$x^\mu$ and~$q$ has been constructed in the
non-relativistic setting in reference~\cite{Linden:1995bt}.
The relativistic ge\-ne\-ralization of this approach proceeds as follows.
(For simplicity, we put the coupling constant~$g$ equal to one.) One introduces a~Lie algebra-valued variable~$\Lambda (\tau)
\equiv \Lambda^a (\tau) T^a$ where the functions~$\Lambda^a (\tau)$ are Grassmann odd, the charge~$q$ being def\/ined as an
expression which is bilinear in~$\Lambda$ (i.e.~a description which is familiar for the spin):
\begin{gather}
q \equiv -\frac{1}2[\Lambda, \Lambda]_+,
\qquad
\text{i.e.}
\qquad
q^a= -\frac{{\rm i}}2f^{abc} \Lambda^b \Lambda^c.
\label{eq:qGrass}
\end{gather}
The Lagrangian
\begin{gather}
L(x, \dot x, \Lambda, \dot \Lambda) \equiv \frac{m}{2}\dot x^2 + \frac{{\rm i}}{2}\Tr\left(\Lambda\frac{D
\Lambda}{d\tau}\right)
= \frac{m}{2}\dot x^2 + \frac{{\rm i}}{2}\Tr(\Lambda \dot \Lambda) + \dot x^\mu\Tr(q A_\mu),
\label{eq:LGrass}
\end{gather}
is invariant under gauge transformations for which $\delta_{\lambda}A_\mu= D_\mu \lambda$ and
\begin{gather*}
%\label{eq:transfL}
\delta_{\lambda} \Lambda (\tau)= -{\rm i}[\Lambda (\tau), \lambda (x (\tau))],
\end{gather*}
which implies the transformation law~\eqref{eq:transfq}.
Moreover, the Lagrangian leads to the following equations of motion for $x^\mu$ and $\Lambda^a$:
\begin{gather}
m \ddot x_\mu= \dot x^\nu\Tr \{q (\partial_\mu A_\nu - \partial_\nu A_\mu) \} + \Tr(\dot q A_\mu),
\qquad
0= \frac{D \Lambda^a}{d\tau}.
\label{eq:eomGrass}
\end{gather}
From~\eqref{eq:qGrass} it follows that $\frac{D q}{d\tau}= - \big[\Lambda, \frac{D \Lambda}{d\tau}\big]_+$, hence~\eqref{eq:eomGrass}
implies that $\frac{D q^a}{d\tau}=0$, i.e.~the charge~$q$ is covariantly conserved.
Substitution of the latter result into the f\/irst of equations~\eqref{eq:eomGrass} yields the equation of motion $m\ddot x^\mu=
\Tr(q F^{\mu\nu})\dot x_\nu $.
We note that the Hamiltonian associated to the Lagrangian~\eqref{eq:LGrass} reads
\begin{gather*}
H= \frac{1}{2m}[p^\mu - \Tr(q A^\mu)] [p_\mu - \Tr(q A_\mu)],
%\label{eq:HGrass}
\end{gather*}
or $H= \frac{m}{2}\dot x^2$ if expressed in terms of the velocity.
The Poisson brackets
\begin{gather*}
\{x^\mu, p_\nu \}= \delta^\mu_\nu,
\qquad
\{\Lambda_a, \Lambda_b \}= -{\rm i} \delta_{ab},
%\label{eq:PGrass}
\end{gather*}
which imply the non-Abelian algebra of charges
\begin{gather*}
\{q_a, q_b \}= f_{abc} q_c,
\end{gather*}
again allow us to recover all previous equations of motion from the evolution equation of dynamical variables~$F$, i.e.~from
$\dot F= \{F, H \}$.
In terms of the kinematical momentum $\Pi_\mu \equiv p_\mu - \Tr(q A_\mu)$, the Hamiltonian reads $H= \frac{1}{2m}
\Pi^2 $ and the Poisson brackets take the form
\begin{gather*}
\{x^\mu, \Pi_\nu \}= \delta^\mu_\nu,
\qquad
\{\Pi_\mu, \Pi_\nu \}= \Tr(q F_{\mu \nu}),
\qquad
\{q_a, q_b \}= f_{abc} q_c.
%\label{eq:PiGrass}
\end{gather*}
The dynamical variable $\Lambda $ which allowed for the Lagrangian formulation is well hidden in the latter equations.

{\bf Continuum formulation of the dynamics.} The f\/ield strength $F_{\mu \nu}$ manifests itself physically by the force f\/ield,
i.e.~by an exchange of energy and momentum between the charge carrier and the f\/ield~\cite{Thirring:1997}.
In the framework of f\/ield theory, the physical entities are described by local f\/ields, i.e.~one has a~continuum formulation.
In order to obtain such a~formulation for the particle's equations of motion~\eqref{wong}, we have to integrate these relations
over the variable~$\tau$ with a~delta function concentrated on the particle's trajectory.
The resulting expressions then involve the current density $j^{a\mu}(y)$ def\/ined in equation~\eqref{JNA} as well as the
energy-momentum tensor (density) of the point particle which is given by (see Appendix~\ref{sec:appendix})
\begin{gather}
\label{eq:energy-momentum}
T^{\mu\nu}(y)=\int d\tau m{\dot x}^\mu{\dot x}^\nu\delta^4  (y-x(\tau) ).
\end{gather}
More explicitly, by using
\begin{gather}
\label{eq:derivedelta}
\dot x^\nu \partial^y_\nu\delta^4  (y-x(\tau) )= - \dot x^\nu \partial^x_\nu\delta^4  (y-x(\tau) )= -
\frac{d}{d \tau}\delta^4  (y-x(\tau) ),
\end{gather}
we have
\begin{gather*}
\partial^y_\nu T^{\nu\mu}(y)= \int d\tau m\dot x^\mu \dot x^\nu \partial^y_\nu\delta^4  (y-x(\tau) )= - \int
d\tau m \dot x^\mu \frac{d}{d \tau} \delta^4  (y-x(\tau) )
\\
\phantom{\partial^y_\nu T^{\nu\mu}(y)}
= \int d\tau m\ddot x^\mu\delta^4  (y-x(\tau) ),
\end{gather*}
and substitution of the particle's equation of motion $ m\ddot x^\mu= \Tr(qF^{\mu \nu}) \dot x_{\nu}$ then yields
\begin{gather*}
\partial^y_\nu T^{\nu\mu}(y)= \int d\tau\dot x_{\nu} q^a F_a^{\mu \nu} (x(\tau))\delta^4  (y-x(\tau) )=
F_a^{\mu \nu} (y)\int d\tau\dot x_{\nu} q^a\delta^4  (y-x(\tau) )
\\
\phantom{\partial^y_\nu T^{\nu\mu}(y)}
= F_a^{\mu \nu} (y)j^a_{\nu} (y).
\end{gather*}
The \emph{continuum version of Wong's equations}~\eqref{wong} thus reads
\begin{gather}
\label{eq:continuumWong}
\partial_\nu T^{\nu\mu}= \Tr(F^{\mu \nu}j_{\nu}),
\qquad
\text{where}
\qquad
D_{\mu} j^{\mu}=0.
\end{gather}
These equations describe the exchange of energy and momentum between the f\/ield $F^{\mu \nu}$ and the current $j^\mu$ (i.e.~the
matter).
They admit an obvious generalization to Moyal space, see equation~\eqref{eq:continuumNCeqs} below.
In Appendix~\ref{sec:appendix}, we show that they also admit a~natural extension to curved space (endowed with a~metric tensor
$(g_{\mu \nu} (x))$).
Moreover, we will prove there that they have to hold for arbitrary dynamical matter f\/ields~$\phi$ whose dynamics is described~by
a~generic action $S[\phi; g_{\mu \nu}, A_{\mu}^a]$ which is invariant under both gauge transformations and general coordinate
transformations ($g_{\mu \nu}$ and $A_{\mu}^a$ representing f\/ixed external f\/ields).
We note that the expectation values of equations~\eqref{eq:continuumWong} viewed as operatorial relations in quantum f\/ield theory
imply Wong's classical equations of motion for suf\/f\/iciently localized, quantum ``wave-packet'' states~\cite{Brown:1979bv}.

{\bf Curved space.} Finally, we also point out that equations which are somewhat similar to Wong's equations appear in general
relativity for a~spinning particle in curved space, for which case the contraction of the Riemann tensor $R^\alpha_{\hphantom{\alpha}\beta\mu\nu}$
with the spin tensor $S^{\mu\nu}$ plays a~role which is similar to the f\/ield strength $F^{\mu\nu}$ in Yang--Mills theories.
The explicit form of these equations of motion, which are known as the \emph{Mathisson--Papapetrou--Dixon
equations}~\cite{Dixon:1970zza, Mathisson:1937zz,Papapetrou:1951pa}, is given~by
\begin{gather*}
\frac{\nabla}{d{\tau}}  (mu^{\alpha} )+\frac{1}2S^{\mu\nu}u^{\sigma} R^{\alpha}_{\hphantom{\alpha}\sigma\mu\nu}=0,
\qquad
\frac{\nabla S^{\alpha\beta}}{d{\tau}}=0,
\end{gather*}
where $\frac{\nabla}{d{\tau}}$ denotes the covariant derivative along the trajectory and $(u^{\mu})$ is the particle's
four-velocity.

\section{Lagrangian approach to particles in NC space}
\label{sec:lagrange}

{\bf Moyal space and distributions.} We consider four dimensional Moyal space, i.e.~we assume that the space-time coordinates
fulf\/ill a~Heisenberg-type algebra (for a~review see~\cite{Blaschke:2010kw,Rivasseau:2007a, Szabo:2001} and references therein).
Thus, the star product of functions is def\/ined~by
\begin{gather*}
 (f \star g  ) (x) \equiv
\big({\rm e}^{\frac{{\rm i}}{2}\theta^{\mu \nu} \partial_\mu^x \partial_\nu^y}f(x) g(y) \big) \big|_{x=y},
\end{gather*}
where the parameters $\theta^{\mu \nu}= - \theta^{\nu \mu}$ are constant, and their star commutator is def\/ined by $\starco{f}{g}
\equiv f \star g - g \star f$, which implies that $\starco{x^\mu}{x^\nu}= {\rm i}\theta^{\mu \nu}$.
In the sequel we will repeatedly use the following fundamental properties of the star product:
\begin{gather}
\int d^4xf \star g= \int d^4xf \cdot g,
\qquad
\int d^4xf \star g \star h= \int d^4xh \star f \star g.
\label{eq:rulesstar}
\end{gather}
An important point to note is that in Moyal space, the integral $\int d^4x$ plays the role of a~trace\footnote{This is best seen
by employing the Weyl map from operators to functions, and is also the reason for the cyclicity property~\eqref{eq:rulesstar}.
Furthermore, when considering gauge f\/ields only the action is gauge invariant, not the Lagrangian.}, and hence equations of
motion must always be derived from the action rather than the Lag\-rangian.
For a~detailed discussion of the algebras of functions and of distributions on Moyal space in the context of non-commutative
spaces and of quantum mechanics in phase space, we refer to~\cite{VignesTourneret:2006xa}
and~\cite{GraciaBondia:1987kw,Varilly:1988jk}, respectively.
Here, we only note that the star product of the delta distribution $\delta_y$ (with support in~$y$) with a~function~$\psi$ may be
def\/ined by application to a~test function~$\varphi$:
\begin{gather*}
\langle \delta_y \star \psi,\varphi \rangle \equiv \int d^4x  (\delta_y \star \psi  ) (x)\varphi(x)=\int
d^4x  (\delta_y \star \psi \star \varphi  ) (x)=\int d^4x\delta_y (x)  (\psi \star \varphi  ) (x)
\\
\phantom{\langle \delta_y \star \psi,\varphi \rangle}
=  (\psi \star \varphi  ) (y).
\end{gather*}
Hence, the action of the distribution $\delta_y \star \psi $ on the test function~$\varphi$ is equal to the action of the
distribution $\delta_y $ on the test function $\psi\star\varphi$.
Similarly, we f\/ind
\begin{gather*}
\langle \psi \star \delta_y,\varphi \rangle \equiv \int d^4x  (\psi \star \delta_y  ) (x)\varphi(x)= \int
d^4x  (\psi \star \delta_y \star \varphi  ) (x)= \int d^4x  (\delta_y \star \varphi \star \psi  ) (x)
\\
\phantom{\langle \psi \star \delta_y,\varphi \rangle}
=  (\varphi \star \psi  ) (y).
\end{gather*}

The following considerations hold for an arbitrary antisymmetric matrix $(\theta^{\mu \nu})$, but for the physical applications
it is preferable to assume that $\theta^{\mu 0}=0$, i.e.~assume the time to be commuting with the spatial coordinates.
This choice is motivated by the fact that the parameters $\theta^{ij}$ have close analogies with a~constant magnetic f\/ield both
from the algebraic and dynamical points of view~\cite{Delduc:2007av}, and by the fact that a~non-commuting time leads to problems
with time-ordering in quantum f\/ield theory~\cite{Bahns:2002}.

{\bf Charged particle in Moyal space.} Since a~$U_\star(1)$ gauge f\/ield $(A^\mu)$ on Moyal space entails a~non-Abelian structure
for the f\/ield strength tensor $(F_{\mu \nu})$ due to the star product~\cite{Blaschke:2010kw,Rivasseau:2007a, Szabo:2001},
\begin{gather*}
F_{\mu \nu} \equiv \partial_\mu A_\nu - \partial_\nu A_\mu - {\rm i} g \starco{A_\mu}{A_\nu},
%\label{eq:Fstar}
\end{gather*}
one expects that the treatment of a~source for this gauge f\/ield as given by a~``point'' particle should allow for a~description
of such particles on non-commutative space which is quite similar to the one found by Wong in the case of Yang--Mills theory.
If the matter content in f\/ield theory is given by a~spinor f\/ield~$\psi$, then the interaction term with the gauge f\/ield reads
\begin{gather*}
\int d^4yJ^\mu \star A_\mu,
\qquad
\text{with}
\qquad
J^\mu \equiv g\bar{\psi} \gamma^\mu \star \psi,
\end{gather*}
i.e.~it involves two star products.
By virtue of the properties~\eqref{eq:rulesstar} one of these star products can be dropped under the integral, but not both of them.
If we consider the particle limit (i.e.~$J^\mu$ representing the current density of the particle), it is judicious to maintain
the star product between $J^\mu$ and $A_\mu$ so as not to hide the non-commutative nature of the underlying space (over which we
integrate) and to allow for the use of the cyclic invariance property~\eqref{eq:rulesstar} later on in our derivation.
In fact, the pairing $\langle J, A\rangle \equiv \int d^4yJ^\mu \star A_\mu$ represents the analogue of the pairing
$\langle j, A\rangle \equiv \int d^4y\Tr(j^\mu A_\mu)$ in Yang--Mills theory on commutative space.
Accordingly, we will require the action for the particle to be invariant under non-commutative gauge transformations def\/ined at
the inf\/initesimal level by $\delta_{\lambda}A_\mu= D_\mu \lambda \equiv \partial_\mu \lambda - {\rm i}g\starco{A_\mu}{\lambda}$
where the parameter~$\lambda$ is an arbitrary function.
As was done for Wong's equations, we assume that the charge~$q$ of the relativistic particle in non-commutative space depends on
the parameter~$\tau$ parametrizing the particle's world line and that it represents a~given non-dynamical variable.
The coupling of this particle to an external $U_\star(1)$ gauge f\/ield $(A^\mu)$ can be described by the action\footnote{We note
that the coupling of $U_{\star} (1)$ gauge f\/ields to external currents has also been addressed in the recent
work~\cite{Adorno:2011wj}.}
\begin{gather}
S[x] \equiv S_{\rm free} [x] - S_{\rm int} [x] \equiv -m \int d\tau\sqrt{\dot{x}^2} - \int d^4yJ^\mu A_\mu,
\label{eq:actNC}
\end{gather}
where
\begin{gather}
J^\mu(y) \equiv \int d\tau q(\tau)\dot x^\mu(\tau)\delta^4 (y-x(\tau) ).
\label{JNC}
\end{gather}

Keeping in mind the formulation of f\/ield theory on non-commutative space~\cite{Blaschke:2010kw,Rivasseau:2007a, Szabo:2001}, we
will argue directly with the action rather than the Lagrangian function and require this action to be invariant under
non-commutative gauge transformations.
For such an inf\/initesimal transformation we have
\begin{gather*}
\delta_\lambda \int d^4yJ^\mu A_\mu=\int d^4yJ^\mu D_\mu\lambda=-\int d^4y(D_\mu J^\mu) \star\lambda.
\end{gather*}
Hence, invariance of the action~\eqref{eq:actNC} under non-commutative gauge transformations requires the current to be
\emph{covariantly conserved,} i.e.~$D_\mu J^\mu=0$.
By virtue of equation~\eqref{JNC} we now infer that
\begin{gather}
0 \stackrel{!}{=}
(D_\mu J^\mu)(y)=\int d\tau q\dot x^\mu D^y_\mu \delta^4(y-x(\tau))
\nonumber
\\
\phantom{0 \stackrel{!}{=}(D_\mu J^\mu)(y)}
=\int d\tau q\dot x^\mu \left\{\partial^y_\mu\delta^4(y-x(\tau))-{\rm i}g\starco{A_\mu(y)}{\delta^4(y-x(\tau))}
\right\}.
\label{eq:q-gauge-cond}
\end{gather}
Equation~\eqref{eq:derivedelta} entails that the f\/irst term in the last line can be rewritten as
\begin{gather*}
\int d\tau q\dot{x}^\mu \partial^y_\mu\delta^4(y-x(\tau))= - \int d\tau q\frac{d}{d\tau} \delta^4(y-x(\tau))= \int
d\tau\frac{dq}{d\tau}\delta^4(y-x(\tau)).
\end{gather*}
From condition~\eqref{eq:q-gauge-cond} it thus follows that the charge~$q$ has to be \emph{covariantly conserved} along the world
line in the sense that
\begin{gather}
0= \int d\tau\frac{Dq}{d\tau}\delta^4(y-x(\tau)) \equiv \int d\tau \left\{\frac{dq}{d\tau}\delta^4(y-x(\tau))
-{\rm i}gq\dot{x}^\mu \starco{A_\mu(y)}{\delta^4(y-x(\tau))} \right\}.
\label{DJNC}
\end{gather}
For later reference, we note that this relation yields the following equality after star multiplication with $A_\nu (y) \delta
y^{\nu} $ and integration over~$y$:
\begin{gather}
 \int d^4y \int d\tau\delta x^{\nu} (\tau) \frac{dq}{d\tau}\delta^4 (y-x(\tau)) \star A_\nu (y)
\nonumber
\\
\qquad
= {\rm i} g \int d^4y \int d\tau\delta x^{\nu} (\tau) q\dot{x}^\mu \starco{A_\mu (y)}{\delta^4 (y-x(\tau))} \star
A_\nu (y)
\nonumber
\\
\qquad
= - {\rm i} g \int d^4y \int d\tau\delta x^{\nu} (\tau) q\dot{x}^\mu \delta^4 (y-x(\tau)) \star \starco{A_\mu
(y)}{A_\nu (y)}.
\label{DJ}
\end{gather}

In order to derive the equation of motion for the particle determined by the action~\eqref{eq:actNC}, we vary the latter with
respect to $x^\mu$.
The variation of $S_{\rm free}$ being the same as in commutative space, we only work out the variation of the interaction part
$S_{\rm int}$, all star products being viewed as functions of the variable~$y$:
\begin{gather*}
\delta S_{\rm int}=\delta \int d^4y \int d\tau\big\{\dot x^\mu q\delta^4(y-x(\tau))\star A_\mu(y) \big\}
\\
\phantom{\delta S_{\rm int}}
=\int d^4y \int d\tau\big\{\delta\dot x^\mu q\delta^4(y-x(\tau))\star A_\mu(y) +\dot x^\mu q\delta
\big[\delta^4(y-x(\tau))\big] \star A_\mu(y)\big\}
\\
\phantom{\delta S_{\rm int}}
=\int d^4y \int d\tau \left\{\frac{d (\delta x^\mu)}{d\tau}q\delta^4(y-x(\tau))\star A_\mu(y)
+\dot x^\mu q(\delta x^\nu)\partial^x_\nu \delta^4(y-x(\tau))\star A_\mu(y)\right\}
\\
\phantom{\delta S_{\rm int}}
=\int d^4y \int d\tau(\delta x^\nu) \left\{- \frac{d}{d\tau} \big[q\delta^4(y-x (\tau))\big]\star A_\nu (y) +\dot
x^\mu q\delta^4(y-x)\star \partial^y_\nu A_\mu(y)\right\}.
\end{gather*}
By virtue of the product rule, the f\/irst term in the last line yields two terms, one involving $\frac{dq}{d\tau}$ which can be
rewritten using relation~\eqref{DJ}, and one involving $\frac{d}{d\tau} \delta^4(y-x(\tau))$ which can be rewritten
using~\eqref{eq:derivedelta}:
\begin{gather*}
 - \int d^4y \int d\tau(\delta x^\nu)q\frac{d}{d\tau} \delta^4(y-x (\tau)) \star A_\nu (y)
\\
\qquad
= \int d^4y \int d\tau(\delta x^\nu)q\dot x^\mu \big[\partial^y_\mu \delta^4(y-x (\tau)) \big] \star A_\nu (y)
\\
\qquad
= -\int d^4y \int d\tau(\delta x^\nu)q\dot x^\mu \delta^4(y-x (\tau)) \star \partial^y_\mu A_\nu (y).
\end{gather*}
Hence, we arrive at
\begin{gather*}
\delta S_{\rm int}=\int d^4y \int d\tau(\delta x^\nu)\dot x^\mu q\delta^4(y-x (\tau)) \star
\big\{\partial^y_\nu A_\mu(y) - \partial^y_\mu A_\nu (y) + {\rm i}g\starco{A_\mu(y)}{A_\nu (y)} \big\}
\\
\phantom{\delta S_{\rm int}}
=\int d^4y \int d\tau(\delta x^\nu)\dot x^\mu q\delta^4(y-x (\tau)) \star F_{\nu\mu}(y)
= \int d\tau(\delta x^\nu)\dot x^\mu qF_{\nu\mu}(x(\tau)).
%\label{eq:viererkraft1}
\end{gather*}
Note that it is the constraint equation~\eqref{DJ} following from~\eqref{DJNC} that yields the terms which are quadratic in the
gauge f\/ield.
This is very much the same mechanism as in the commutative space calculation which leads to Wong's equations,
cf.~\eqref{eq:point-action-comm-un}--\eqref{wong}.

In conclusion, we obtain the following {\em equation of motion for the charged relativistic particle in non-commutative space:}
\begin{gather}
m\ddot x^\mu=q F^{\mu\nu}\dot x_\nu,
\qquad
\text{with}
\qquad
F_{\mu \nu} \equiv \partial_\mu A_\nu - \partial_\nu A_\mu - {\rm i} g \starco{A_\mu}{A_\nu}.
\label{eq:NCeom}
\end{gather}
We note that the antisymmetry of $F_{\mu \nu}$ with respect to its indices implies
\begin{gather*}
0= \ddot x^\mu \dot x_\mu= \frac{1}2\frac{d \dot x^2}{d\tau},
\end{gather*}
which is consistent with $\dot x^2=1$.
Consistency of the equation of motion~\eqref{eq:NCeom} requires its gauge invariance, i.e.~the gauge invariance of its
right-hand-side.
Since
\begin{gather*}
\delta_{\lambda} (q F^{\mu\nu}\dot x_\nu)= q(\delta_{\lambda} F^{\mu\nu}) \dot x_\nu= - {\rm i} gq \starco{F^{\mu\nu}}{\lambda}
\dot x_\nu= gq \theta^{\rho \sigma} (\partial_{\rho} F^{\mu \nu}) (\partial_{\sigma} \lambda) \dot x_\nu + {\cal O} \big(\theta^2\big),
\end{gather*}
\textit{the gauge invariance only holds for constant field strengths} by contrast to the case of Wong's equation for Yang--Mills
f\/ields in commutative space.
This dif\/ference can be traced back to the fact that the analogue of the trace in Yang--Mills theory is given in Moyal space~by
the integral $\int d^4x$: since such an integral does not occur in the dif\/ferential equation~\eqref{eq:NCeom}, the gauge
invariance is only realized for constant $F_{\mu \nu}$.
We will further discuss the latter f\/ields and the comparison with Yang--Mills theory in~Section~\ref{sec:constfield} where we
will also consider the subsidiary condition for the charge~$q$.
Here we only note the following points in this respect.
The equation of motion~\eqref{eq:NCeom} taken for itself is consistent for any constant values of~$q$ and $F_{\mu \nu}$.
Furthermore in non-commutative space the auxiliary variable~$q$ cannot be rendered dynamical with a~non-trivial gauge
transformation law~\eqref{eq:transfq} for $q(\tau)$ since the star commutator of $q(\tau)$ with a~gauge transformation parameter
$\lambda (x(\tau))$ vanishes.

In summary, the coupling of a~relativistic particle to a~gauge f\/ield $(A^\mu)$ is described in general by the Lagrangian
\begin{gather*}
L(x, \dot x)= -m \sqrt{\dot x^2} - q A_\mu \dot x^\mu,
\qquad
\text{or}
\qquad
L(x, \dot x)= \frac m2\dot x^2 + q A_\mu \dot x^\mu.
%\label{eq:LinLag}
\end{gather*}
The non-commutativity of space-time can be implemented in the Lagrangian framework by rewriting the interaction term of the
action as an integral $\int d^4y (J^\mu\star A_\mu ) (y)$ where the current~$J^\mu$ is def\/ined by~\eqref{JNC} and
by requiring this action to be invariant under non-commutative gauge transformations.
The resulting equation of motion~\eqref{eq:NCeom} is only gauge invariant for constant f\/ield strengths.

{\bf Continuum formulation of the dynamics.} By following the same lines of arguments as for the Lorentz--Yang--Mills force
equation (see equations~\eqref{eq:energy-momentum}--\eqref{eq:continuumWong}), we can obtain a~continuum version of the equation
of motion~\eqref{eq:NCeom} by multiplying this equation with $\delta^4 (y-x(\tau)) \varphi(y)$, where $\varphi(y)$ is
a~suitable test function, and integrating over~$\tau$ and over~$y$.
More explicitly, by starting from the energy-momentum tensor~\eqref{eq:energy-momentum} for the point particle and using
relation~\eqref{eq:derivedelta}, we get
\begin{gather*}
\int d^4y(\partial_\nu T^{\nu\mu})(y) \varphi (y)= \int d^4y\int d\tau m\dot x^\mu \dot x^\nu \partial^y_\nu\delta^4
(y-x(\tau)) \varphi(y)
\\
\qquad
= - \int d^4y\int d\tau m \dot x^\mu \frac{d}{d \tau} \delta^4 (y-x(\tau)) \varphi(y)
= \int d^4y\int d\tau m\ddot x^\mu\delta^4 (y-x(\tau)) \varphi(y).
\end{gather*}
Substitution of equation~\eqref{eq:NCeom} then yields the expression
\begin{gather*}
\int d^4y\int d\tau q F^{\mu\nu} (x(\tau)) \dot x_\nu\delta^4 (y-x(\tau)) \varphi(y)= \int d^4y\int d\tau q
F^{\mu\nu} (y) \dot x_\nu\delta^4 (y-x(\tau)) \varphi(y)
\\
\qquad
=\int d^4y(F^{\mu\nu} J_\nu)(y)\varphi (y),
\end{gather*}
where we considered the current density~\eqref{JNC} in the last line.
Thus, we have the result
\begin{gather}
\label{eq:continuumNCeqs}
\partial_\nu T^{\nu\mu}= F^{\mu\nu} J_\nu,
\qquad
\text{where}
\qquad
D_{\mu} J^{\mu}=0,
\end{gather}
and these relations are completely analogous to the continuum equations~\eqref{eq:continuumWong} which correspond to Wong's
equations (apart from the fact that the invariance of~\eqref{eq:continuumNCeqs} under non-commutative gauge transformations
requires the f\/ield strength $F^{\mu \nu}$ to be constant).

\section{Case of a~constant f\/ield strength}
\label{sec:constfield}

We will now discuss the dynamics of charged particles coupled to a~constant f\/ield strength on the basis of the results obtained
in Sections~\ref{sec:recallcommspace} and~\ref{sec:lagrange}.
Indeed this case represents a~mathematically tractable and physically interesting application of the general formalism.

We successively discuss the case of non-Abelian Yang--Mills theory on Minkowski space and the case of a~$U_{\star}(1)$ gauge
f\/ield on Moyal space while emphasizing the dif\/ferences that exist for constant f\/ield strengths.

\subsection{Wong's equations in commutative space}

The case of a~``uniform f\/ield strength'' in Yang--Mills theory has been addressed some time ago by the authors of
reference~\cite{Brown:1979bv}, see also~\cite{Estienne:2011fn} for related points.
Since the Yang--Mills f\/ield strength $F_{\mu \nu} (x) \equiv F_{\mu \nu}^a (x) T^a$ is not gauge invariant, but transforms under
f\/inite gauge transformations as $F_{\mu \nu}^{\prime}= U^{-1} F_{\mu \nu} U$ (with $U(x) \in G=$ structure group), one has
to specify f\/irst what is meant by a~constant f\/ield.
The f\/ield $F_{\mu \nu}$ is said to be \textit{uniform} if the gauge f\/ield $A_{\mu} (x) \equiv A_{\mu}^a (x) T^a$ at a~point~$x$
can be related by a~gauge transformation to the gauge f\/ield $A_{\mu} (y)$ at any other point~$y$.
More precisely, for a~space-time translation parametrized by $a \in {\mathbb{R}}^4$, there exists a~gauge transformation
$x\mapsto U(x; a) \in G$ such that
\begin{gather*}
A_{\mu} (x +a)= U^{-1} (x; a) A_{\mu} (x) U(x; a) + {\rm i}U^{-1} (x; a) \partial_{\mu} U (x; a),
\end{gather*}
and thereby
\begin{gather*}
F_{\mu \nu} (x + a)= U^{-1} (x; a) F_{\mu \nu} (x) U(x; a).
\end{gather*}
In this case a~gauge may be chosen in which all components of $F_{\mu \nu}$ are constant because $F_{\mu \nu} $ at the point~$x$
can be made equal to its value at some arbitrary point~$y$ by transforming it by an appropriate gauge group element.

Since the f\/ield strength $F_{\mu \nu} \equiv \partial_{\mu} A_{\nu} - \partial_{\nu} A_{\mu} + {\rm i} [A_{\mu}, A_{\nu}]$
contains two terms, namely $ \partial_{\mu} A_{\nu} - \partial_{\nu} A_{\mu}$ which has the Abelian form, and $ [A_{\mu},
A_{\nu}]$ which does not involve derivatives, a~constant non-zero f\/ield strength $F_{\mu \nu}$ can be obtained either from
a~linear gauge potential (i.e.~an Abelian-like gauge f\/ield),
\begin{gather*}
A_{\mu}= - \frac{1}{2}F_{\mu \nu} x^{\nu},
\qquad
\partial_{\mu} A_{\nu} - \partial_{\nu} A_{\mu}= F_{\mu \nu}= \const,
\qquad
[A_{\mu}, A_{\nu}]=0,
\end{gather*}
or from a~constant non-Abelian-like gauge potential,
\begin{gather*}
A_{\mu}= \const,
\qquad
\partial_{\mu} A_{\nu}=0,
\qquad
{\rm i}[A_{\mu}, A_{\nu}]= F_{\mu \nu}= \const.
\end{gather*}
In fact~\cite{Brown:1979bv}, these two types of potentials exhaust all possibilities for a~constant f\/ield strength.
It has been shown for the structure group $\text{SU}(2)$ that the two types of gauge potentials leading to a~same constant f\/ield
strength are gauge inequivalent and result in physically dif\/ferent behavior when matter interacts with them, e.g.\
the solutions of Wong's equations have completely dif\/ferent properties in both cases.
An explicit example for $G= \text{SU}(2)$ is given by a~constant magnetic f\/ield in~$z$-direction~\cite{Estienne:2011fn}: let
$\sigma_k$ (with $k=1,2,3$) denote the Pauli matrices and suppose
\begin{gather*}
F_{0i}=0,
\qquad
F_{ij}= \varepsilon_{ijk} B_k,
\qquad
\text{with}
\qquad
(B_k)_{k=1,2,3}= (0,0, 2 \sigma_3).
\end{gather*}
This constant f\/ield strength derives from the linear Abelian-like potential
\begin{gather*}
\vec A\equiv (A_k)_{k=1,2,3}= - \frac{1}{2}\vec x \wedge \vec B= (-y, x, 0)\sigma_3,
\end{gather*}
or from a~constant non-Abelian-like potential $\vec A= (-\sigma_2, \sigma_1, 0)$.

\subsection{Wong's equations in non-commutative space}

The f\/ield strength associated to a~$U_{\star}(1)$ gauge f\/ield $(A_{\mu})$ on Moyal space reads
\begin{gather*}
F_{\mu \nu} \equiv \partial_{\mu} A_{\nu} - \partial_{\nu} A_{\mu} - {\rm i} g \starco{A_{\mu}}{A_{\nu}}= \partial_{\mu} A_{\nu}
- \partial_{\nu} A_{\mu} + g \theta^{\rho \sigma} (\partial_{\rho} A_{\mu})(\partial_{\sigma} A_{\nu}) + {\cal O}\big(\theta^2\big).
\end{gather*}
Due to the derivatives appearing in the star commutator, the Abelian-like term $\partial_{\mu} A_{\nu} - \partial_{\nu} A_{\mu}$
and the non-Abelian-like term $- {\rm i} g \starco{A_{\mu}}{A_{\nu}}$ cannot vanish independently of each other: the
non-commutative f\/ield strength $F_{\mu \nu}$ can only be constant for a~linear Abelian-like potential.
More precisely, for
\begin{gather}
\label{linGF}
A_\mu= -\frac12\bar{B}_{\mu \nu} x^\nu,
\end{gather}
where the coef\/f\/icients $\bar{B}_{\mu \nu} \equiv - \bar{B}_{\nu \mu}$ are constant, we obtain
\begin{gather}
F_{\mu \nu}= \bar{B}_{\mu \nu} - \frac{g}{4}\bar{B}_{\mu \rho} \theta^{\rho \sigma} \bar{B}_{\sigma \nu}.
\label{eq:NCFS}
\end{gather}
This f\/ield strength is constant, but dependent on the non-commutativity parameters $\theta^{\mu \nu}$.
If we interpret $\bar{B}_{\mu \nu}= \partial_\mu A_\nu - \partial_\nu A_\mu $ as the physical f\/ield strength,
then~\eqref{eq:NCFS} means that the non-commutativity parameters $\theta^{\mu \nu}$ modify in general the trajectories of the
particle as compared to its motion in commutative space\footnote{We note that the latter dependence on the non-commutativity
parameters can be eliminated mathematically if one assumes that $\bar{B}_{\mu \nu}$ depends in a~specif\/ic way on the parameters
$\theta^{\mu \nu}$ and some~$\theta$-independent constants ${B}_{\mu \nu}$.
To illustrate this point~\cite{Delduc:2007av}, we assume that the only non-vanishing components of $ \bar{B}_{\mu \nu}$ and
$\theta^{\mu \nu}$ are as follows: $\bar{B}_{12}= - \bar{B}_{21} \equiv \bar B, \theta^{12}= - \theta^{21} \equiv \theta$,
i.e.~$F_{12}= \bar B (1 + \frac{g}{4}\theta \bar B)= - F_{21}$.
If $\bar B$ depends on~$\theta$ and on a~$\theta$-independent constant~$B$ according to
\begin{gather*}
\bar{B}= \bar{B} (B; \theta) \equiv \frac2{g \theta}\big(\sqrt{1 + g \theta B}-1\big)
=B\left(1-\frac{g}{4}\theta B\right)+{\cal O}\big(\theta^2\big),
\end{gather*}
then relation~\eqref{eq:NCFS} implies that the non-commutative f\/ield strength $ F_{12} $ is a~$\theta$-indepen\-dent constant: $
F_{12}= B$.}.

Since the f\/ield strength transforms under a~f\/inite gauge transformation $U(x) \in U_{\star}(1)$ as $F_{\mu \nu}= U^{-1} \star
F_{\mu \nu} \star U$, a~constant f\/ield $F_{\mu \nu}$ is gauge invariant.
Thus, for constant f\/ield strengths the situation is quite dif\/ferent for $ U_{\star}(1)$ gauge f\/ields and for non-Abelian gauge
f\/ields on Minkowski space despite the fact that we encounter the same structure for the gauge transformations, the f\/ield strength
and the action functional in both cases.

Let us again come back to the expression~\eqref{linGF} for the gauge f\/ield.
Substitution of this expression into the subsidiary condition~\eqref{DJNC} yields
\begin{gather}
0= \int d\tau \big\{\dot q\delta^4(y-x(\tau)) - {\rm i}gq\dot{x}^\mu
\starco{A_\mu(y)}{\delta^4(y-x(\tau))} \big\}
\nonumber
\\
\phantom{0}
= \int d\tau \big\{\dot q\delta^4(y-x(\tau)) - \frac{g}{2}q\dot{x}^\mu \bar{B}_{\mu \rho} \theta^{\rho \sigma}
\partial^y_{\sigma} \delta^4(y-x(\tau)) \big\}.
\label{SubsidCond}
\end{gather}
If the matrix $(\bar{B}_{\mu \nu})$ is the inverse of the matrix $(\theta^{\mu \nu})$,
i.e.~$\bar{B}_{\mu \rho} \theta^{\rho\sigma}= \delta^{\sigma}_{\mu}$, then $F_{\mu \nu}= (1 - \frac{g}{4})\bar{B}_{\mu \nu}$ and, by virtue
of~\eqref{eq:derivedelta} and an integration by parts, condition~\eqref{SubsidCond} takes the form
\begin{gather*}
0= \int d\tau\dot q\delta^4(y-x(\tau)) \left\{1 - \frac{g}{2} \right\}.
\end{gather*}
The latter relation is obviously satisf\/ied for a~constant~$q$.
In this case, the equation of motion~\eqref{eq:NCeom} for the particle in non-commutative space, i.e.~$m \ddot{x}^\mu= q F^{\mu\nu} \dot{x}_\nu$,
has the same form as the one of an electrically charged particle in ordinary space.
This result is analogous to the one obtained for a~constant magnetic f\/ield in $x^3$-direction within the Hamiltonian approaches,
see equations~\eqref{eq:standeom} and~\eqref{eq:strangeconst} below.

\section{Hamiltonian approaches to particles in NC space}
\label{sec:ham}

To start with, we brief\/ly review the Hamiltonian approaches in commutative space before considering the generalization to the
non-commutative setting.
In the latter setting, we will notice that various approaches yield dif\/ferent results since several expressions which coincide in
commutative space no longer agree.

\subsection{Reminder on the Poisson bracket approach}

The Hamiltonian formulation of relativistic (as well as non-relativistic) mechanics is based on two inputs (e.g.\
see reference~\cite{Marsden:1999}): a~Hamiltonian function and a~Poisson structure (or equivalently a~symplectic structure).
If one starts from the Lagrangian formulation, the Hamiltonian function is obtained from the Lagrange function by a~Legendre
transformation.
E.g.\
the Lagrangian $L (x, \dot x)= \frac m2\dot x^2 + e A_\mu \dot x^\mu$ (involving the constant charge~$e$) yields the
\emph{Hamiltonian}
\begin{gather}
H (x,p)= \frac1{2m}(p-eA)^2= \frac1{2m}(p^\mu-eA^\mu) (p_\mu-eA_\mu).
\label{eq:Hamilton}
\end{gather}
The trajectories in phase space are parametrized by $\tau \mapsto (x (\tau), p (\tau) ) $ where~$\tau$ denotes a~real
variable to be identif\/ied with proper time after the equations of motion have been derived.
The \emph{Poisson brackets} $\{\cdot, \cdot \}$ of the phase space variables $x^\mu$, $p^\mu$ are chosen in such a~way that the
evolution equation $\dot F= \{F, H \}$ (where $\dot F \equiv d F /d\tau$) yields the Lagrangian equation of motion for~$x^\mu$,
though written as a~system of f\/irst order dif\/ferential equations.
For instance, if we consider the usual form of the Poisson brackets, i.e.\ the \emph{canonical Poisson brackets}
\begin{gather}
\{x^\mu, x^\nu \}= 0,
\qquad
\{p^\mu, p^\nu \}= 0,
\qquad
\{x^\mu, p^\nu \}= \eta^{\mu \nu},
\label{eq:uxxpp}
\end{gather}
then substitution of $F= x^\mu$ and $F= p^\mu$ into $\dot F= \{F, H \}$ (with~$H$ given by~\eqref{eq:Hamilton}) yields the
system of equations
\begin{gather*}
m \dot x^\mu= p^\mu-eA^\mu,
\qquad
m \dot p^\mu= e (p_\nu-eA_\nu) \partial^{\mu} A^\nu,
%\label{eq:ucaneqs}
\end{gather*}
from which we conclude that
\begin{gather*}
m \ddot x^\mu= \dot p^\mu-e \dot A^\mu= \underbrace{\frac em(p_\nu-eA_\nu)}_{=e\dot x_\nu} \partial^\mu A^\nu - e
\dot x^\nu \partial_\nu A^\mu= e \dot x_\nu\big(\partial^\mu A^\nu - \partial^\nu A^\mu \big) \equiv ef^{\mu \nu} \dot x_\nu.
\end{gather*}
This equation coincides with the Euler--Lagrange equation for $x^\mu$ following from
the Lagrangian $L (x, \dot x)= \frac m2\dot x^2 + e A_\mu \dot x^\mu$.

Concerning the gauge invariance, we emphasize a~result~\cite{Thirring:1997} which does not seem to be very well known.
The Hamiltonian~\eqref{eq:Hamilton} is \emph{not} invariant under a~gauge transformation $A_\mu \to A_\mu + \partial_\mu \lambda$
which is quite intriguing.
However, it is invariant if this transformation is combined with the phase space transformation $(x^\mu, p^\mu) \to (x^\mu, p^\mu
+ e \partial^\mu \lambda)$, the latter being a~canonical transformation since it preserves the fundamental Poisson
brackets~\eqref{eq:uxxpp}.
Indeed, under this combined transformation the kinematical momentum $\Pi^\mu \equiv p^\mu -e A^\mu$ (which coincides with~$m \dot
x^\mu$) is invariant.

We note that the Hamiltonian~\eqref{eq:Hamilton} can be rewritten in terms of the variable $\Pi_\mu \equiv p_\mu - eA_\mu$ as $H
\equiv \frac1{2m}\Pi^2 $.
Thereby~$H$ has the form of a~free particle Hamiltonian, but the Poisson brackets are now modif\/ied: from $\Pi_\mu= p_\mu -
eA_\mu$ and~\eqref{eq:uxxpp} it follows that
\begin{gather}
\{x^\mu, x^\nu \}= 0,
\qquad
\{\Pi^\mu, \Pi^\nu \}= e f^{\mu \nu} (x),
\qquad
\{x^\mu, \Pi^\nu \}= \eta^{\mu \nu},
\label{eq:uxxpipi}
\end{gather}
with $f^{\mu \nu} \equiv \partial^\mu A^\nu - \partial^\nu A^\mu$.
Since the electromagnetic f\/ield strength $f^{\mu \nu}$ is gauge invariant, the latter invariance is manifestly realized in this
formulation.

In summary, the coupling of a~charged particle to an electromagnetic f\/ield can either be described by the canonical Poisson
brackets~\eqref{eq:uxxpp} and the minimally coupled Hamiltonian~\eqref{eq:Hamilton} or by introducing the f\/ield strength into the
Poisson brackets (as a~non-commutativity of the momenta) and considering a~Hamiltonian which has the form of a~free particle
Hamiltonian.

\subsection{Reminder on the symplectic form approach}

If we gather all phase space variables into a~vector $\vec{\xi} \equiv ({\xi}^I) \equiv (x^0, \dots,x^3, p^0, \dots, p^3)$, the
fundamental Poisson brackets~\eqref{eq:uxxpp} read
\begin{gather*}
\big\{\xi^I, \xi^J \big\}= \Omega^{IJ},
\qquad
\text{with}
\qquad
\big(\Omega^{IJ}\big) \equiv
\begin{bmatrix}
0 & \eta^{\mu \nu}
\\
- \eta^{\mu \nu} & 0
\end{bmatrix}.
%\label{eq:xixi}
\end{gather*}
The inverse of the Poisson matrix $\Omega \equiv (\Omega^{IJ})$ is the matrix with entries $\omega_{IJ} \equiv
(\Omega^{-1})_{IJ}$ which def\/ines the \emph{symplectic $2$-form}
\begin{gather}
\label{eq:symplform}
\omega \equiv \frac12\sum\limits_{I,J}\omega_{IJ}d{\xi}^I \wedge d{\xi}^J= dp^\mu \wedge dx_\mu,
\end{gather}
e.g.\
see reference~\cite{Marsden:1999} for mathematical details.
The Hamiltonian equations of motion can be written as
\begin{gather*}
\dot{\xi}^I= \big\{\xi^I, H \big\}= \Omega^{KJ}\partial_K \xi^I \partial_J H,
\qquad
\text{i.e.}
\qquad
\dot{\xi}^I= \Omega^{IJ}\partial_J H,
\end{gather*}
or equivalently as $\omega_{IJ}\dot{\xi}^J= \partial_I H $.

In terms of the phase space variables $(x^\mu, \Pi^\mu)$ appearing in the non-canonical Poisson brac\-kets~\eqref{eq:uxxpipi}, the
symplectic $2$-form~\eqref{eq:symplform} reads
\begin{gather*}
\omega= d\Pi^\mu \wedge dx_\mu + \frac12e f_{\mu \nu}dx^\mu \wedge dx^\nu.
\end{gather*}
This formulation of the electromagnetic interaction based on the symplectic $2$-form
and the evolution equation $\omega_{IJ}\dot{\xi}^J= \partial_I H $ goes back to the seminal work of Souriau~\cite{Souriau:1997}.

\subsection{Standard (Poisson bracket) approach to NC space-time}
%\label{sec:poiss}

In order to introduce a~non-commutativity for the conf\/iguration space, one generally starts from a~function~$H$ on phase space to
which one refers as the Hamiltonian without any reference to a~Lag\-rangian, e.g.\
we can consider the function~$H$ given in equation~\eqref{eq:Hamilton}.
The non-commutativity of space-time is then implemented by virtue of the Poisson brackets
\begin{gather}
\{x^\mu, x^\nu \}= \theta^{\mu \nu},
\qquad
\{p^\mu, p^\nu \}= 0,
\qquad
\{x^\mu, p^\nu \}= \eta^{\mu \nu},
\label{eq:xxpp}
\end{gather}
where $\theta^{\mu \nu}= - \theta^{\nu \mu}$ is again assumed to be constant.
(For an overview of the description of non-relativistic charged particles in non-commutative space we refer
to~\cite{Acatrinei:2011zz, Delduc:2007av,Horvathy:2010wv}, the pioneering work being~\cite{Duval:2000xr,Duval:2001hu}, see
also~\cite{Acatrinei:2001xs,Acatrinei:2002si, Mezincescu:2000zq,Nair:2000ii} for some subsequent early work.
We also mention that dynamical systems in non-commutative space can be constructed by applying Dirac's treatment of constrained
Hamiltonian systems to an appropriate action functional, see~\cite{Deriglazov:2002xc} and references therein.)

As in the commutative setting, we gather all phase space variables into a~vector $\vec{\xi} \equiv ({\xi}^I) \equiv (x^0,
\dots,x^3, p^0, \dots, p^3)$, the fundamental Poisson brackets~\eqref{eq:xxpp} being now given~by
\begin{gather*}
\{\xi^I, \xi^J \}= \Omega^{IJ},
\qquad
\text{with}
\qquad
(\Omega^{IJ}) \equiv
\begin{bmatrix}
\theta^{\mu \nu} & \eta^{\mu \nu}
\\
- \eta^{\mu \nu} & 0
\end{bmatrix}.
\end{gather*}
Quite generally, the Poisson bracket of two arbitrary functions $F$, $G$ on phase space reads
\begin{gather*}
\{F, G \} \equiv \sum\limits_{I,J} \Omega^{IJ}\partial_I F\partial_J G= \theta^{\mu \nu}\frac{\partial F}{\partial
{x}^\mu}\frac{\partial G}{\partial {x}^\nu} + \frac{\partial F}{\partial {x}^\mu}\frac{\partial G}{\partial {p}_\mu} -
\frac{\partial F}{\partial {p}^\mu}\frac{\partial G}{\partial {x}_\mu}.
%\label{eq:PoissFG}
\end{gather*}

Substitution of $F= x^\mu$ and $F= p^\mu$ into $\dot F= \{F, H \}$ (with~$H$ given by~\eqref{eq:Hamilton}) yields the system of
equations
\begin{gather}
m \dot x^\mu= (p_\nu-eA_\nu) (\eta^{\mu \nu} - e \theta^{\mu \rho} \partial_{\rho} A^\nu),
\nonumber
\\
m \dot p^\mu= e (p_\nu-eA_\nu) \partial^{\mu} A^\nu.
\label{eq:caneqs}
\end{gather}
In the present case, the phase space transformation $(x^\mu, p^\mu) \to (x^\mu, p^\mu + e \partial^\mu \lambda)$ does not
represent a~canonical transformation since it does not preserve the Poisson brackets~\eqref{eq:xxpp} if $\theta^{\mu \nu} \neq0$.
Hence, the resulting Hamiltonian equations of motion~\eqref{eq:caneqs} are not gauge invariant, as has already been pointed out
in reference~\cite{Delduc:2007av} by considering dif\/ferent gauges.

In the next two subsections, we recall how this problem can be overcome for the particular case of a~constant f\/ield strength as
well as more generally, and we compare with the results obtained in Section~\ref{sec:lagrange} from the action involving star
products.
Here, we only note that a~non-Abelian structure of the f\/ield strength is hidden in equation~\eqref{eq:caneqs}.
To illustrate this point, we consider the particular case where the only non-vanishing components of $ \theta^{\mu \nu}$ are
$\theta^{ij}= \varepsilon^{ij} \theta$ (with $i,j \in \{1, 2 \}$ and $\varepsilon^{12} \equiv -\varepsilon^{21} \equiv 1$) and
where the only non-vanishing components of $A^\mu$ are $A^i (x^1, x^2)$ (with $i \in \{1, 2 \}$).
For this situation which describes a~time-independent magnetic f\/ield perpendicular to the $x^1 x^2$-plane, the f\/irst of
equations~\eqref{eq:caneqs} yields
\begin{gather*}
m \dot x_i= (p_k-eA_k) (\delta_{ik} - e \theta \varepsilon_{ij} \partial_{j} A_k),
\end{gather*}
and implies
\begin{gather}
m\frac{d}{dt}  (x_i + e \theta \varepsilon_{ij} A_j  )= (1 + e{\cal F}_{12})(p_i -eA_i)
\label{eq:NCHeom}
\end{gather}
with
\begin{gather}
{\cal F}_{12} \equiv \partial_1 A_2 - \partial_2 A_1 +e\{A_1, A_2 \}= \partial_1 A_2 - \partial_2 A_1 +e\theta^{\rho
\sigma} (\partial_{\rho} A_1)(\partial_{\sigma} A_2).
\label{eq:fmn}
\end{gather}
Thus, we f\/ind a~non-Abelian structure for the generalized f\/ield strength, but in the present approach the f\/ield ${\cal F}_{\mu
\nu} $ is only linear in the non-commutativity parameters in contrast to the f\/ield $F_{\mu \nu} $ in~\eqref{eq:NCeom} which
involves the star commutator
\begin{gather*}
- {\rm i} \starco{A_\mu}{A_\nu}= \theta^{\rho \sigma} (\partial_{\rho} A_\mu)(\partial_{\sigma} A_\nu) +{\cal O} \big(\theta^2\big).
\end{gather*}
If the gauge potential is linear in~$x$, the f\/ield strengths ${\cal F}_{\mu \nu}$ and $F_{\mu \nu}$ as def\/ined~by
equations~\eqref{eq:fmn} and~\eqref{eq:NCeom}, respectively, coincide with each other (if one identif\/ies the coupling
constant~$g$ with~$e$).

To conclude, we note that~\eqref{eq:NCHeom} can be solved for $p_i-eA_i$ in terms of $m \dot x_i$: the system of f\/irst order
dif\/ferential equations~\eqref{eq:caneqs} can then be written as a~second order equation for $x^\mu$, but the resulting equations
of motion are not gauge invariant and they cannot be derived from a~Lagrangian~\cite{Acatrinei:2011zz}.

\subsection{Standard approach to NC space-time continued}
%\label{sec:standardapproachcontinued}

The reasoning presented concerning the brackets~\eqref{eq:uxxpipi} suggests to consider a~Hamiltonian which has a~free form and
to introduce a~f\/ield strength $B^{\mu \nu} (x)$ as a~non-commutativity of the momenta, i.e.~consider phase space variables
$(x^\mu, p^\mu)$ satisfying the non-canonical Poisson algebra
\begin{gather}
\{x^\mu, x^\nu \}= \theta^{\mu \nu},
\qquad
\{p^\mu, p^\nu \}= e B^{\mu \nu},
\qquad
\{x^\mu, p^\nu \}= \eta^{\mu \nu},
\label{eq:utxxpipi}
\end{gather}
with $\theta^{\mu \nu}$ constant.
As pointed out in reference~\cite{Horvathy:2002wc}, the Jacobi identities for the algebra~\eqref{eq:utxxpipi} are only satisf\/ied
if \emph{the field strength is constant:}
\begin{gather*}
\{x^\mu, \{p^\nu, p^\lambda \} \} + \{\text{cyclic permutations of $\mu$, $\nu$, $\lambda$}\}=e \theta^{\mu \rho}
\partial_{\rho} B^{\nu \lambda}.
\end{gather*}
Thus, the dynamics of a~charged particle coupled to a~\emph{constant field} $B^{\mu \nu}$ on non-commutative space-time can be
described in terms of phase space variables $(x^\mu, p^\mu)$ satisfying the non-canonical Poisson algebra~\eqref{eq:utxxpipi},
the Hamiltonian being given by $H (p)= \frac1{2m}p^2= \frac1{2m}p^\mu p_\mu$.
The Hamiltonian equations of motion
\begin{gather*}
m\dot x^\mu= p^\mu,
\qquad
m\dot p^\mu= e B^{\mu \nu} p_\nu,
\end{gather*}
then imply the second order equation
\begin{gather}
\label{eq:standeom}
m \ddot x^\mu= e B^{\mu \nu}\dot x_\nu.
\end{gather}
This equation of motion for $x^{\mu}$ coincides with the one that one encounters for $\theta^{\mu \nu}=0$ since the Hamiltonian
only depends on~$p$ and not on the coordinates $x^\mu$ whose Poisson brackets do not vanish.
However the non-commutativity parameters $\theta^{\mu \nu}$ appear in quantities like the volume form on phase space which is the
$4$-fold exterior product of the symplectic form with itself,
\begin{gather*}
%\label{volform}
dV \equiv \frac{1}{4!}\omega^4= \frac{1}{\sqrt{\text{det}\Omega}} d\xi^1 \cdots d\xi^{8},
\end{gather*}
where~$\Omega$ denotes the Poisson matrix and where we suppressed the exterior product symbol.

\subsection{\texorpdfstring{``Exotic''}
{Exotic} (symplectic form) approach to NC space-time}

The Hamiltonian approach to mechanics on non-commutative space based on the simple \linebreak form~\eqref{eq:utxxpipi} of the Poisson
algebra (in which the Poisson bracket $\{x^\mu, p^\nu \}$ has the canonical form) has been nicknamed the \emph{standard
approach}.
As we just recalled, it does not allow for the inclusion of a~non-constant f\/ield strength.
By contrast, the so-called \emph{exotic approach}~\cite{Duval:2001hu,Horvathy:2010wv} which is based on a~simple form of the
symplectic $2$-form allows us to describe generic f\/ield strengths~$ B_{\mu \nu} (\vec x)$.
In this setting, the constant non-commutativity parameters $\theta^{\mu \nu}$ are introduced into the symplectic
$2$-form\footnote{One may as well consider $\vec p$-dependent parameters $\theta^{\mu \nu}$.} def\/ined on the phase space
parametrized by $(x^\mu, p^\mu)$:
\begin{gather*}
\omega= dp^\mu \wedge dx_\mu + \frac12e B_{\mu \nu}dx^\mu \wedge dx^\nu + \frac12\theta_{\mu \nu}dp^\mu \wedge dp^\nu.
%\label{eq:2form}
\end{gather*}
The Poisson matrix is obtained by the inversion of the symplectic matrix (e.g.\
see reference~\cite{Vanhecke:2005rq} for the case of a~space-time of arbitrary dimension), and therefore it has a~more
complicated form than the one corresponding to~\eqref{eq:utxxpipi}.
By way of illustration, we recall the result that one obtains for the simplest instance~\cite{Duval:2001hu} where one has only
two spatial coordinates, i.e.~$\vec x \equiv (x_1, x_2)$:
\begin{gather*}
\{{x}_1, {x}_2 \}= \kappa^{-1}\theta,
\qquad
\{{p}_1, {p}_2 \}= \kappa^{-1}e B,
\qquad
\{{x}_i, {p}_j \}= \kappa^{-1}\delta_{ij},
%\label{eq:2dexotic}
\end{gather*}
where $\kappa (\vec x) \equiv 1 - e \theta B (\vec x) $ with $\theta_{12} \equiv \theta$ and $B_{12} \equiv B$.
None of the brackets now has a~canonical form.
The equations of motion following from the Hamiltonian $H (\vec x, \vec p) \equiv \frac{1}{2m}\vec p^{2} + e V (\vec x) $ read
\begin{gather}
\dot p_i= e E_i + e B \varepsilon_{ij} \dot x_j,
\qquad
\text{with}
\qquad
E_i \equiv - \partial_i V,
\qquad
i\in \{1, 2 \}
\nonumber
\\
m^{\ast} \dot x_i= p_i - e m \theta \varepsilon_{ij} E_j,
\qquad
\text{with}
\qquad
m^{\ast}\equiv \kappa m,
\qquad
\kappa (\vec x)= 1 - e \theta B (\vec x),
\label{eq:exoeom}
\end{gather}
where $\varepsilon_{ij}$ denotes the components of the constant antisymmetric tensor normalized by $\varepsilon_{12}=1$.
The parameter $m^{\ast}(\vec x) \equiv m\kappa (\vec x)$ may be viewed as an ef\/fective mass depending on the position
of the particle.
Various physical applications of this system of evolution equations have been found in recent years, see~\cite{Horvathy:2010wv}
and references therein.
For $V \equiv 0$, we have $ p_i= m^{\ast} \dot x_i=m \kappa \dot x_i$, hence
\begin{gather*}
\dot p_i= m \dot \kappa \dot x_i + m \kappa \ddot x_i,
\qquad
\text{with}
\qquad
\dot \kappa= -e \theta \dot B= - e \theta \dot x_j \partial_j B.
\end{gather*}
Substitution of this expression into the f\/irst of equations~\eqref{eq:exoeom} yields a~second order dif\/ferential equation for
$x_i$:
\begin{gather}
\label{eq:strange}
m^{\ast} (\vec x)\ddot x_i= e \varepsilon_{ij} \dot x_j B^{\ast},
\qquad
\text{with}
\qquad
B^{\ast} \equiv B + \frac{1}{2}m \theta\varepsilon_{ij} \dot x_i (\partial_j B).
\end{gather}
This equation, which looks somewhat exotic, includes a~$\theta$-dependent term depending on the derivative of the f\/ield strength
and it involves an $\vec x$-dependent mass, i.e.~there is a~dependence of parameters on the localization of the particle in the
space in which it evolves.

The expressions in~\eqref{eq:strange} simplify greatly in the case of a~\emph{constant magnetic field:}
equation~\eqref{eq:strange} then reduces to
\begin{gather}
\label{eq:strangeconst}
m \ddot x_i= e\frac{B}{\kappa}\varepsilon_{ij} \dot x_j,
\qquad
\text{with}
\qquad
\kappa= 1 - e \theta B=\const.
\end{gather}
As was pointed out earlier~\cite{Horvathy:2006nf}, this equation of motion coincides with the ``standard approach''
equation~\eqref{eq:standeom} after a~rescaling of time $t \to \kappa t$.
We note that the value $\check B \equiv \frac{B}{1 - e \theta B}$ coincides with the one obtained for a~constant magnetic f\/ield
in two dimensions from the Seiberg--Witten map in non-commutative gauge f\/ield theory~\cite{Delduc:2007av}, but it dif\/fers from the
constant non-commutative f\/ield strength
\begin{gather*}
F_{12}\equiv \partial_1 A_2 - \partial_2 A_1 - {\rm i} e \starco{A_1}{A_2}= \partial_1 A_2 - \partial_2 A_1 +e\{A_1, A_2\},
\end{gather*}
e.g.\
in the symmetric gauge $(A_1, A_2)= (- \frac{B}{2}x_2, \frac{B}{2}x_1)$, where one f\/inds $F_{12}= B (1 + \frac{e \theta
B}{4 \kappa})$.

In conclusion, dif\/ferent Hamiltonian formulations for a~charged ``point'' particle in a~non-commutative space lead to dif\/ferent
results.
However, for the special case of a~constant magnetic f\/ield strength we have seen in the previous two subsections that the
dif\/ferent Hamiltonian formulations lead to the same results (or to results that are related to each other by a~redef\/inition of
the magnetic f\/ield).
So does the Lagrangian formulation of Section~\ref{sec:lagrange} as we have shown in Section~\ref{sec:constfield}.

\section{Concluding remarks}
Just as there exist dif\/ferent approaches to the formulation of gauge f\/ield theories on non-commutative spaces (e.g.\
the star product approach~\cite{Szabo:2001}, the approach of spectral triples~\cite{Chamseddine:2006ep}, of matrix
models~\cite{DuboisViolette:1988vq}, \ldots), there appear to exist dif\/ferent approaches to the dynamics of relativistic or
non-relativistic particles in non-commutative space which are subject to a~background gauge f\/ield.
It is plausible that these approaches yield essentially the same results in the particular case of a~constant magnetic f\/ield,
i.e.~a f\/ield strength which does not depend on the non-commuting coordinates.
The ``exotic'' (symplectic form) approach to non-commutative space-time can be viewed as an extension of all other approaches to
the case of a~generic f\/ield strength.

\appendix
\section{Continuum formulation on a~generic manifold}
\label{sec:appendix}

In this appendix, we show that Wong's equations, as formulated on a~generic space-time manifold, admit a~simple continuum
version.
Moreover, we will prove that the latter formulation has to hold for arbitrary dynamical matter f\/ields~$\phi$ whose dynamics is
described by a~generic action $S[\phi; g_{\mu \nu}, A_{\mu}^a]$ which is invariant under both gauge transformations and general
coordinate transformations ($g_{\mu \nu}$ and $A_{\mu}^a$ representing f\/ixed external f\/ields).
These arguments generalize to Moyal space in the particular case of a~constant f\/ield strength.

Let~$M$ be a~four dimensional space-time manifold endowed with a~f\/ixed metric tensor $(g_{\mu \nu})$ of signature $(+,-,-,-)$.
We denote the covariant derivative of a~tensor f\/ield with respect to the Levi-Civita-connection by $\nabla_{\mu}$ (e.g.\
$\nabla_{\mu} V^{\nu}= \partial_{\mu} V^{\nu} + \Gamma^{\nu}_{\mu \rho} V^{\rho}$ where the coef\/f\/icients $ \Gamma^{\nu}_{\mu
\rho}$ are the Christof\/fel symbols) and the gauge covariant derivative as before by $D_{\mu}$ (e.g.\
$\delta_{\lambda} A_{\mu}^a= D_{\mu} \lambda^a \equiv \partial_{\mu} \lambda^a - {\rm i} g [A_{\mu}, \lambda]^a$ for the
inf\/initesimal gauge transformation of the Yang--Mills gauge f\/ield $(A_{\mu}^a)$).
Since we used the notation $\frac{Dq^a}{d\tau} \equiv \dot x^{\mu} D_{\mu} q^a$ in the main body of the text, we will write
$\frac{\nabla V^{\mu}}{d\tau} \equiv \dot x^{\nu} \nabla_{\nu} V^{\mu}$ for the derivative of the vector f\/ield $V^{\mu} (x
(\tau))$ along the trajectory $\tau \mapsto x(\tau)$.

{\bf Lorentz-force and its non-Abelian generalization.} The Lorentz-force equation on the space-time manifold~$M$ reads
\begin{gather}
m\frac{\nabla u^{\mu}}{d\tau}=e F^{\mu}_{\hphantom{\mu}\nu} u^{\nu},
\label{eq:Lor}
\end{gather}
where $u^{\mu} \equiv \dot{x}^{\mu}$ denotes the $4$-velocity of the particle of constant charge $q\equiv e$ and where $F_{\mu
\nu}$ represents a~given electromagnetic f\/ield strength.
This equation of motion follows from the point particle action
\begin{gather*}
S[x]=\frac{m}{2}\int d\tau g_{\mu \nu} (x(\tau))\dot{x}^{\mu}\dot{x}^{\nu} +e\int d\tau A_{\mu} (x(\tau))\dot{x}^{\mu}
%\label{eq:actparticle}
\end{gather*}
upon variation with respect to $x^{\mu}$.

The natural generalization of~\eqref{eq:Lor} to non-Abelian Yang--Mills theory is given by \emph{Wong's equations} as written
\emph{on the space-time manifold~$M$:}
\begin{gather}
m\frac{\nabla^2 x^{\mu}}{d\tau^2}=q^{a} F^{a \mu}_{\hphantom{a \mu}\nu}\dot{x}^{\nu},
\qquad
\text{where}
\qquad
\frac{D q^{a}}{d\tau}=0.
\label{eq:Lor-YM}
\end{gather}
Here, the covariant constancy of the charge-vector $(q^{a})$ represents the geometrically natural generalization of the ordinary
constancy of the charge~$e$ appearing in the Abelian gauge theory.
The equation of motion of $x^{\mu}$ follows from the action functional
\begin{gather}
S_{\rm W} [x]=\frac{m}{2}\int d\tau g_{\mu \nu} (x(\tau))\dot{x}^{\mu}\dot{x}^{\nu} +\int d\tau q^a A^a_{\mu} (x(\tau))
\dot{x}^{\mu}.
\label{eq:Wactparticle}
\end{gather}

{\bf Continuum formulation.} The components ${T}^{\mu \nu}$ of the energy-momentum tensor (density) and the components of the
current density may be def\/ined as functional derivatives of the action,
\begin{gather*}
{T}^{\mu \nu}(x) \equiv 2\frac{\delta S_{\rm W}}{\delta g_{\mu \nu}(x)},
\qquad
{j}_{a}^{\mu}(x) \equiv \frac{\delta S_{\rm W}}{\delta A_{\mu}^{a}(x)}
\end{gather*}
so that expression~\eqref{eq:Wactparticle} implies
\begin{gather*}
{T}^{\mu \nu}(y)= \int d\tau{\delta}^{4}(y-x(\tau))m \dot{x}^{\mu}(\tau)\dot{x}^{\nu}(\tau),
\\
{j}^{a\mu}(y)= \int d\tau{\delta}^{4}(y-x(\tau))q^{a}(\tau)\dot{x}^{\mu}(\tau).
%\label{eq:EM_YM_cur}
\end{gather*}
We note that the energy-momentum $4$-vector is then given by $P^{\mu}= \int_{{\mathbb{R}}^3} d^3 xT^{\mu 0}$ which yields
the standard expressions:
\begin{gather*}
P^{0}= \int_{{\mathbb{R}}^3} d^3 xT^{0 0}= m\dot x^0= m\frac{dt}{d\tau}= \frac{m}{\sqrt{1 - \vec v\,{}^{2}}},
\qquad
P^{i}= m\dot x^i= \frac{mv^i}{\sqrt{1 - \vec v\,{}^{2}}}.
\end{gather*}

The $4$-divergence of the energy-momentum tensor can be evaluated by substituting the equation of motion $m\frac{\nabla^2
x^{\mu}}{d\tau^2}=q^{a} F^{a \mu}_{\hphantom{a \mu}\nu} \dot{x}^{\nu}$:
\begin{gather*}
\nabla_{\mu} {T}^{\mu \nu}(y)= \int d\tau(\dot{x}^{\mu}\nabla^y_{\mu}){\delta}^{4}(y-x(\tau)) m\dot{x}^{\nu}(\tau)
= \int d\tau m\frac{\nabla^{2}x^{\nu}}{d\tau^{2}}(\tau){\delta}^{4}(y-x(\tau))
\\
\phantom{\nabla_{\mu} {T}^{\mu \nu}(y)}
= \int d\tau q^{a}(\tau)F^{a\nu}_{\hphantom{a\nu}\mu} (x(\tau))\dot{x}^{\mu}(\tau){\delta}^{4}(y-x(\tau))
= F^{a\nu}_{\hphantom{a\nu}\mu} (y){j}^{a\mu}(y).
\end{gather*}
Similarly, substitution of the charge transport equation $\frac{Dq^{a}}{d\tau}=0$ into the gauge covariant divergence of the
current density gives
\begin{gather*}
D_{\mu} {j}^{a \mu}(y)= \int d\tau(\dot{x}^{\mu}D^y_{\mu}){\delta}^{4}(y-x(\tau))q^{a}(\tau)
=\int d\tau{\delta}^{4}(y-x(\tau))\frac{Dq^{a}(\tau)}{d\tau}= 0.
\end{gather*}
Therefore the continuum version of equations~\eqref{eq:Lor-YM} reads
\begin{gather}
\nabla_{\nu} {T}^{\nu \mu}= F^{a\mu}_{\hphantom{a\mu}\nu} {j}^{a\nu},
\qquad
\text{where}
\qquad
D_{\mu} {j}^{a\mu}= 0.
\label{eq:cont_Lor_YM}
\end{gather}
These relations may be called \textit{continuum Lorentz--Yang--Mills equations}.

{\bf General derivation of the continuum equations.} Actually equations~\eqref{eq:cont_Lor_YM} do not only hold for point
particles but in a~rather general context as will be shown in the sequel.
To this end let us consider an arbitrary action functional
\begin{gather*}
S=S[\phi; g_{\mu \nu},A_{\mu}^{a}],
\end{gather*}
where $(g_{\mu \nu})$ and $(A_{\mu}^{a})$ denote a~f\/ixed $4$-geometry and Yang--Mills potential respectively, whereas~$\phi$
denotes arbitrary dynamical matter f\/ields.
Taking the action~$S$ to be gauge invariant entails the vanishing of its gauge variation:
\begin{gather*}
0= \delta_{\lambda}S
=\int\left(\frac{\delta S}{\delta\phi}\delta_{\lambda}\phi+\frac{\delta S}{\delta A_{\mu}^{a}}\delta_{\lambda}A_{\mu}^{a}\right).
\end{gather*}
Together with the matter f\/ield equations of motion $\delta S/\delta\phi=0$ and the gauge variation of the Yang--Mills connection,
$\delta_{\lambda}A_{\mu}^{a}=D_{\mu}\lambda^{a}$, this implies
\begin{gather}
D_{\mu} {j}_{a}^{\mu}=0,
\qquad
\text{where}
\qquad
{j}_{a}^{\mu}(x) \equiv \frac{\delta S}{\delta A_{\mu}^{a}(x)},
\label{eq:cur-cons}
\end{gather}
i.e.~the second of equations~\eqref{eq:cont_Lor_YM}.

The fact that~$S$ is geometrically well def\/ined is ref\/lected by its invariance under general coordinate transformations
(dif\/feomorphisms).
The latter are generated by a~generic vector f\/ield $\xi \equiv \xi^{\mu} \partial_{\mu}$.
Thus, we have
\begin{gather}
0= \delta_{\xi}S=\int\left(\frac{\delta S}{\delta\phi}\delta_{\xi}\phi+\frac{\delta S}{\delta g_{\mu \nu}}\delta_{\xi}g_{\mu
\nu}+\frac{\delta S}{\delta A_{\mu}^{a}}\delta_{\xi}A_{\mu}^{a}\right),
\label{eq:action_diff_inv}
\end{gather}
where the matter f\/ield equations again imply the vanishing of the f\/irst term.
The metric tensor f\/ield and the Yang--Mills connection $1$-form $A \equiv A_{\mu} dx^{\mu} \equiv A_{\mu}^a T^a dx^{\mu}$
transform~\cite{Thirring:1997} with the Lie derivative with respect to the vector f\/ield $\xi$:
\begin{gather}
\delta_{\xi}g_{\mu \nu}= (L_{\xi}g)_{\mu \nu}= \nabla_{\mu}\xi_{\nu}+\nabla_{\nu}\xi_{\mu},
\nonumber
\\
\delta_{\xi}A_{\mu}=(L_{\xi}A)_{\mu} \equiv ((i_{\xi} d+d i_{\xi}) A)_{\mu}
= (i_{\xi} (dA- {\rm i}\frac{g}{2}[A,A]) - {\rm i} g [A, i_{\xi} A] + d i_{\xi} A)_{\mu}
\nonumber
\\
\phantom{\delta_{\xi}A_{\mu}}
= \xi^{\nu}F_{\nu \mu}+D_{\mu}(\xi^{\nu}A_{\nu}).
\label{eq:diffeo-var}
\end{gather}
Here, $i_{\xi}$ denotes the inner product of dif\/ferential forms with the vector f\/ield $\xi$
and $F \equiv dA- {\rm i}\frac{g}{2}[A,A] \equiv \frac{1}{2} F_{\mu \nu} dx^{\mu} \wedge dx^{\nu} $ the Yang--Mills curvature $2$-form.
Substitution of the variations~\eqref{eq:diffeo-var} into~\eqref{eq:action_diff_inv} and use of relation~\eqref{eq:cur-cons} now
yields
\begin{gather*}
\nabla_{\mu} {T}^{\mu \nu}= F^{a\nu}_{\hphantom{a\nu}\mu}{j}_{a}^{\mu},
\qquad
\text{where}
\qquad
{T}^{\mu \nu}(x) \equiv 2\frac{\delta S}{\delta g_{\mu \nu}(x)},
\end{gather*}
i.e.~the f\/irst of equations~\eqref{eq:cont_Lor_YM}, thereby completing the proof of our claim.

\subsection*{Acknowledgements}

D.B.~is a~recipient of an APART fellowship of the Austrian Academy of Sciences, and is also grateful for the hospitality of the
theory division of LANL and its partial f\/inancial support.
F.G.~wishes to thank Fabien Vignes-Tourneret for a~useful discussion on the Moyal algebra.
We wish to thank the anonymous referees as well as the editors for their pertinent and constructive comments which contributed to
the clarif\/ication of several points, as well as for pointing out several relevant references.

\pdfbookmark[1]{References}{ref}
\LastPageEnding

\end{document}